\documentclass{appolb}
\usepackage{epsfig}
\def\ev{\,{\rm eV}}
\def\gev{\,{\rm GeV}}
\def\lessim{\mathrel{\rlap{\raise.5ex\hbox{$<$}}{\lower.5ex\hbox{$\sim$}}}}
\def\gtrsim{\mathrel{\rlap{\raise.5ex\hbox{$>$}}{\lower.5ex\hbox{$\sim$}}}}

\def\c{{\rm c}}

\def\data{{\rm data}}

\def\GZK{{\rm GZK}}

\def\MC{{\rm MC}}
\def\min{{\rm min}}

\begin{document}
\pagestyle{plain}
\newcount\eLiNe\eLiNe=\inputlineno\advance\eLiNe by -1
\title{New physics from ultrahigh energy cosmic rays%
}
\author{Subir Sarkar
\address{Department of Physics, Oxford University, 1 Keble Road, 
         Oxford OX1 3NP, UK}}
\maketitle

\begin{abstract}
Observations of cosmic rays with energies above
$\sim4\times10^{10}\gev$ have inspired several speculative suggestions
concerning their origin. The crucial question is whether or not the
spectrum exhibits the expected `GZK cutoff' at this energy ---
concerning which there are presently contradictory results. If there
is indeed a cutoff, then the sources are cosmologically distant and
rather exotic in nature. If there is no cutoff then new physics is
required.
\end{abstract}

\section{Introduction}

As we approach the centennial of the discovery of cosmic radiation by
Victor Hess in 1912, the astrophysical origin of these high energy
particles remains still unknown. Even so the study of cosmic rays has
been extremely rewarding for particle physics. As Hess noted in his
Nobel Lecture in 1936: {\em ``The investigation of \ldots cosmic rays
\ldots has led to the discovery of the positron \ldots It is likely
that further research into ``showers'' and ``bursts'' of the cosmic
rays may possibly lead to the discovery of still more elementary
particles \ldots of which the existence has been postulated by some
theoretical physicists in recent years''}. Indeed several such
discoveries were made in rapid succession --- the muon in 1937, the
pion in 1947 and strange particles (kaons, hyperons) also in
1947. However with the development of particle accelerators in the
1950's, the attention of particle physicists turned naturally to
controlled laboratory experiments. No other new particles have
subsequently been discovered in cosmic rays although there have been
several false alarms (e.g. free quarks, monopoles, \ldots) and many
claims of unexplained phenomena (e.g. `Centauro' events). Today we
have another such mystery --- the observation of ultra high energy
cosmic rays (UHECRs) which cannot propagate very far through the
cosmic microwave background, are unlikely to be significantly affected
by cosmic magnetic fields, and yet which do not point back to any
plausible astrophysical sources in our vicinity
\cite{Nagano:ve,Anchordoqui:2002hs}. The intriguing question is
whether an explanation is possible in terms of known physics or will
require physics beyond the Standard Model.

\section{Observational Status}

\begin{figure}[htb]
\centering
\includegraphics[height=0.4\textwidth]{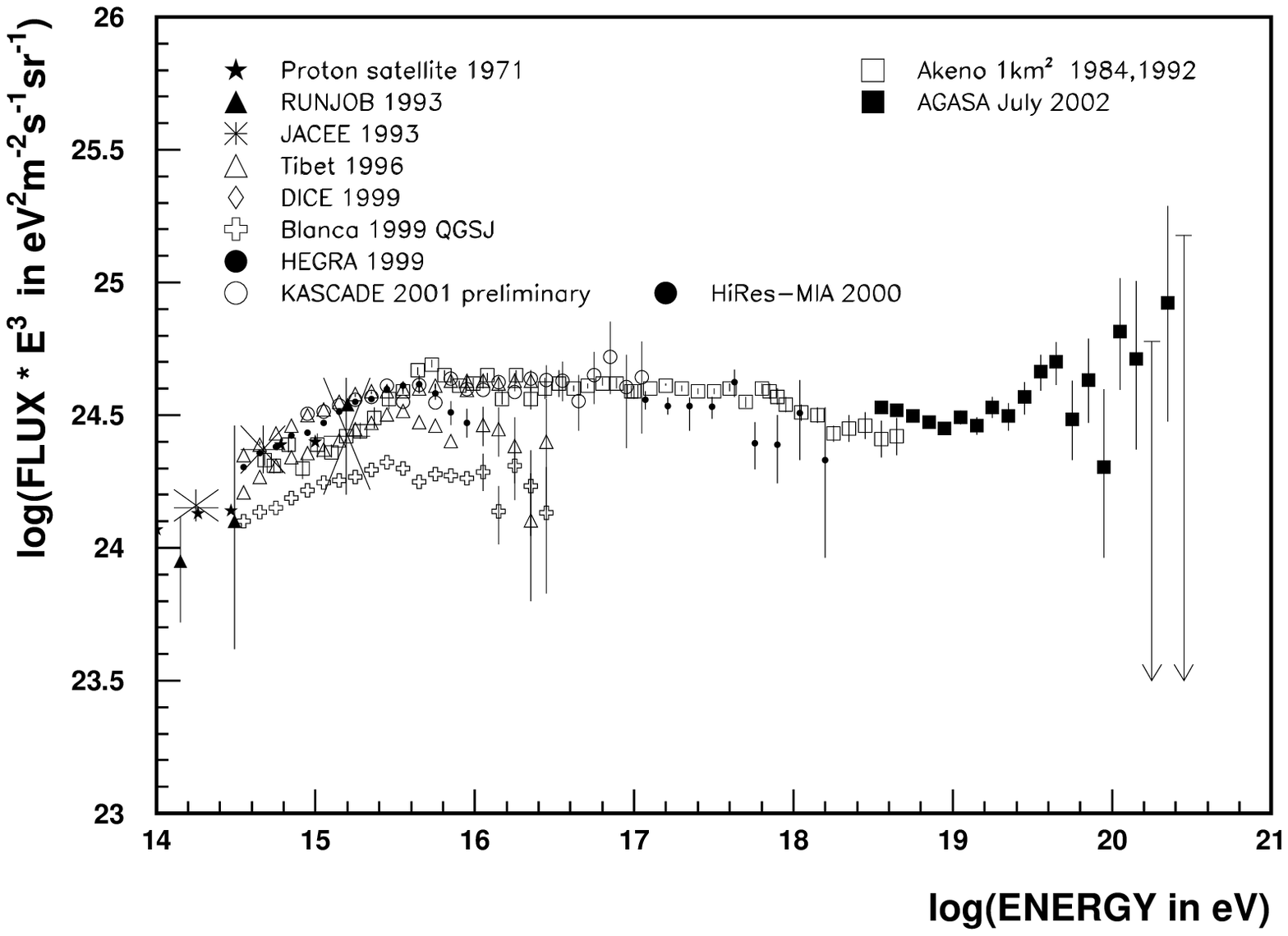}
\includegraphics[height=0.45\textwidth]{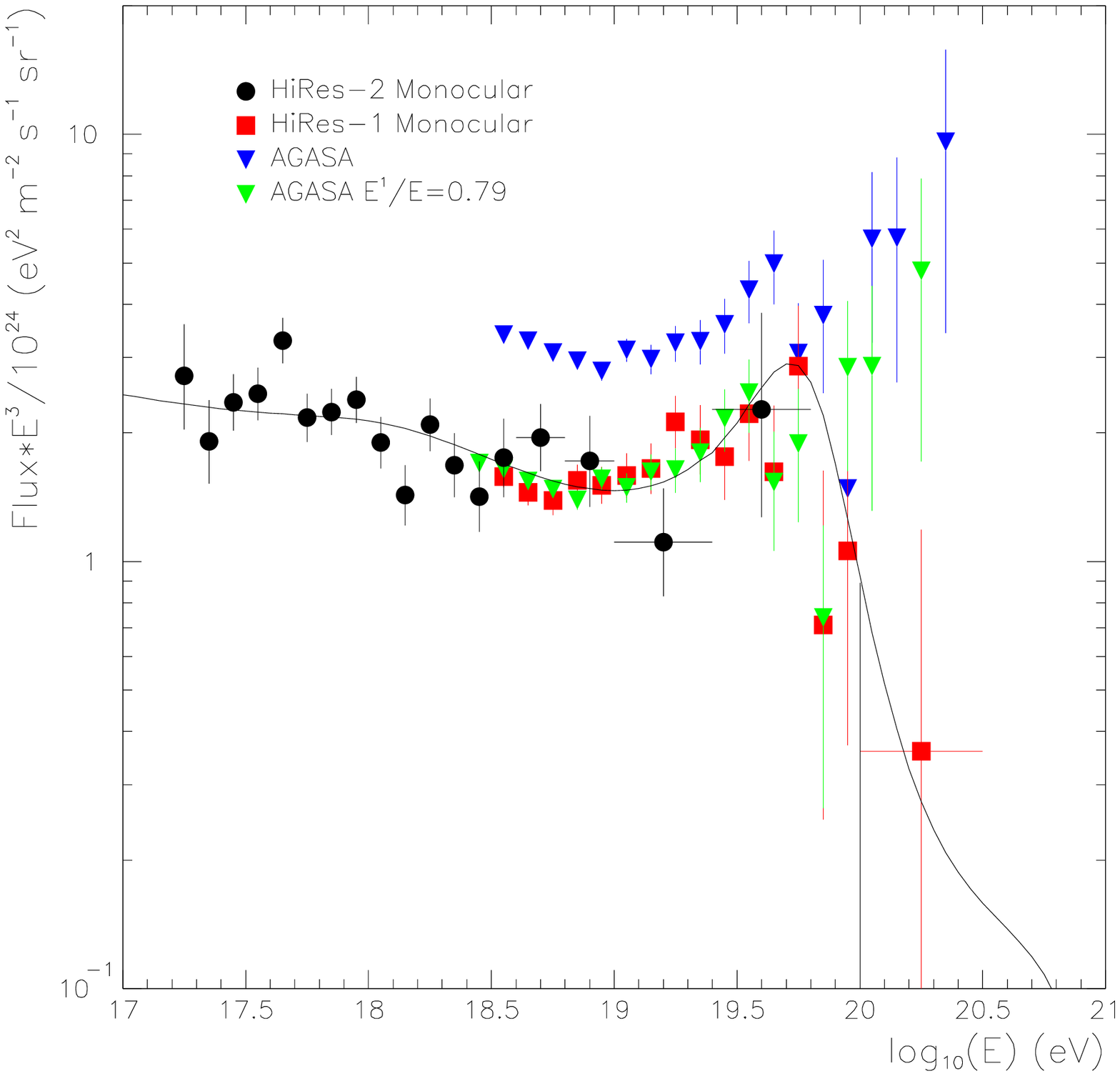}
\caption{The left panel shows the recalibrated AGASA spectrum along
with data from other experiments at lower energies
\protect\cite{Takeda:2002at}. The right panel shows the HiRes data
(fitted to a two-component source model incorporating a GZK cutoff for
the extragalactic component); the AGASA data are shown for comparison,
and with the event energies reduced by 20\%
\protect\cite{Bergman:2002pw}. (All fluxes have been multiplied by
$E^3$).}
\label{spec}
\end{figure}

As we have just heard from Prof Teshima, the AGASA collaboration has
reevaluated the energy determination method used for analysing their
data on air showers collected over 12 years, with particular attention
to the lateral distribution, attenuation with zenith angle, shower
front structure, delayed particles observed far from the shower core,
etc \cite{Takeda:2002at}. They confirm that the energies assigned to
AGASA events have an event-reconstruction accuracy of $\pm 25\%$ at
$10^{11}\gev$, while the systematic uncertainty is $\pm 18\%$
(independent of primary energy above $10^{10}\gev$). As seen in
Fig.~\ref{spec}, the AGASA data at the highest energies connects
smoothly with the (better sampled) Akeno data at lower energies. There
are 59 events with $E>E_\GZK\simeq4\times10^{10}\gev$ (and 8 events
beyond $10^{11}\gev$) which have a spectrum distinctly flatter than at
lower energies, suggestive of a different origin, but with no
indication of the GZK cutoff expected if the sources are
cosmologically distant.

By contrast the HIRES air fluorescence experiment, with a similar
exposure in the mono mode, has reported only 1 event above
$10^{11}\gev$, consistent with a GZK cutoff
\cite{Abu-Zayyad:2002sf}. But as is clear from Fig.~\ref{spec}, the
absolute fluxes are lower than those measured by AGASA and the
begining of the `ankle' in the spectrum is distinctly lower,
suggestive of an energy calibration mismatch bewtween the fluorescence
and air shower detector data. (Since the spectrum falls as
$\sim\,E^{-3}$ below the ankle, the fractional error in the flux is
twice that in the energy, keeping in mind the change in the
differential energy interval with the energy.) If the AGASA energies
are lowered by $\sim20\%$, the spectral shapes can be matched below
$10^{11}\gev$ \cite{Bergman:2002pw}, however 5 AGASA events 
still remain above this energy. Given the low event statistics, the
significance of the discrepancy between the two experiments is not
overwhelming, nevertheless it is clearly of paramount importance to
resolve the issue and establish a consistent energy scale. The
engineering array of the Pierre Auger Observatory \cite{Blumer:2003bc}
has already collected several `hybrid' events observed using both type
of detectors so progress is expected soon.

\begin{figure}[htb]
\centering
\includegraphics[width=0.5\textwidth]{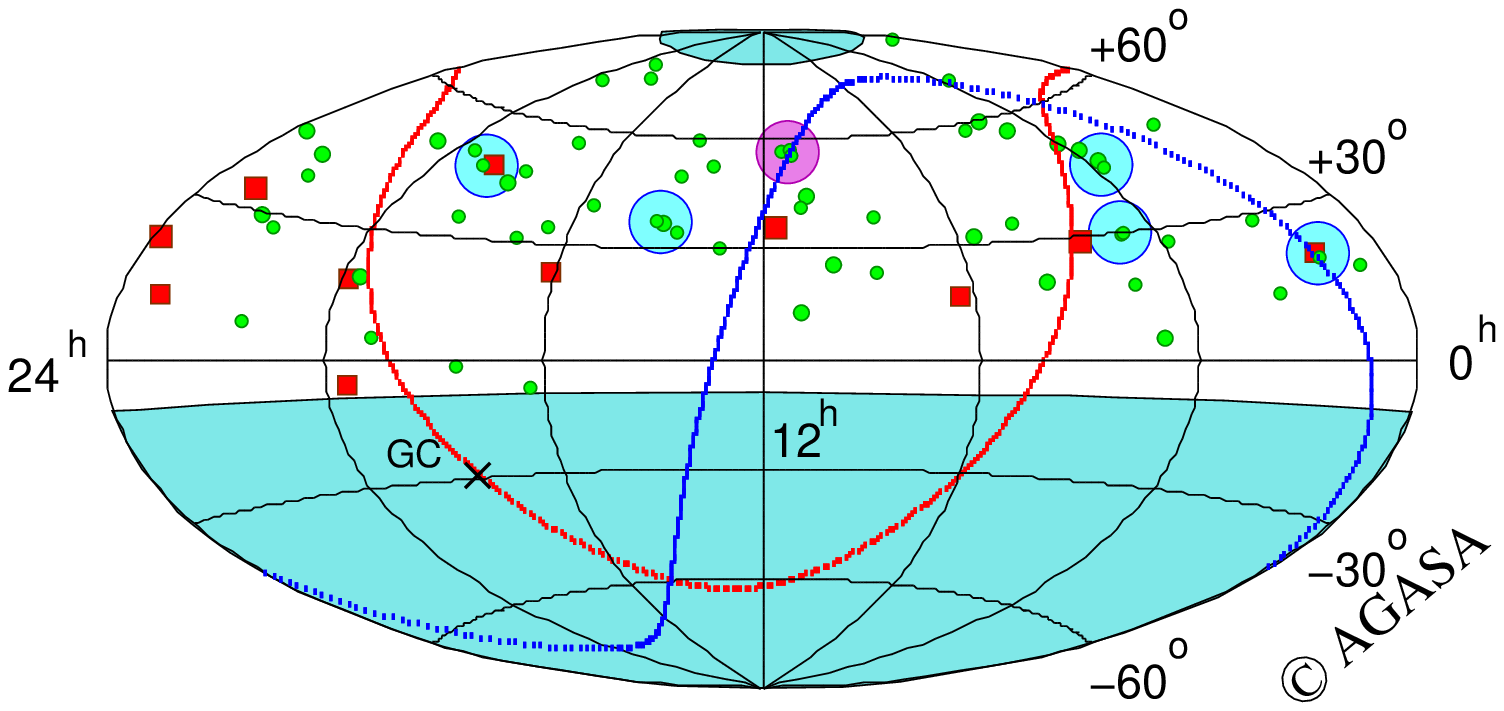}\\
\includegraphics[width=0.45\textwidth]{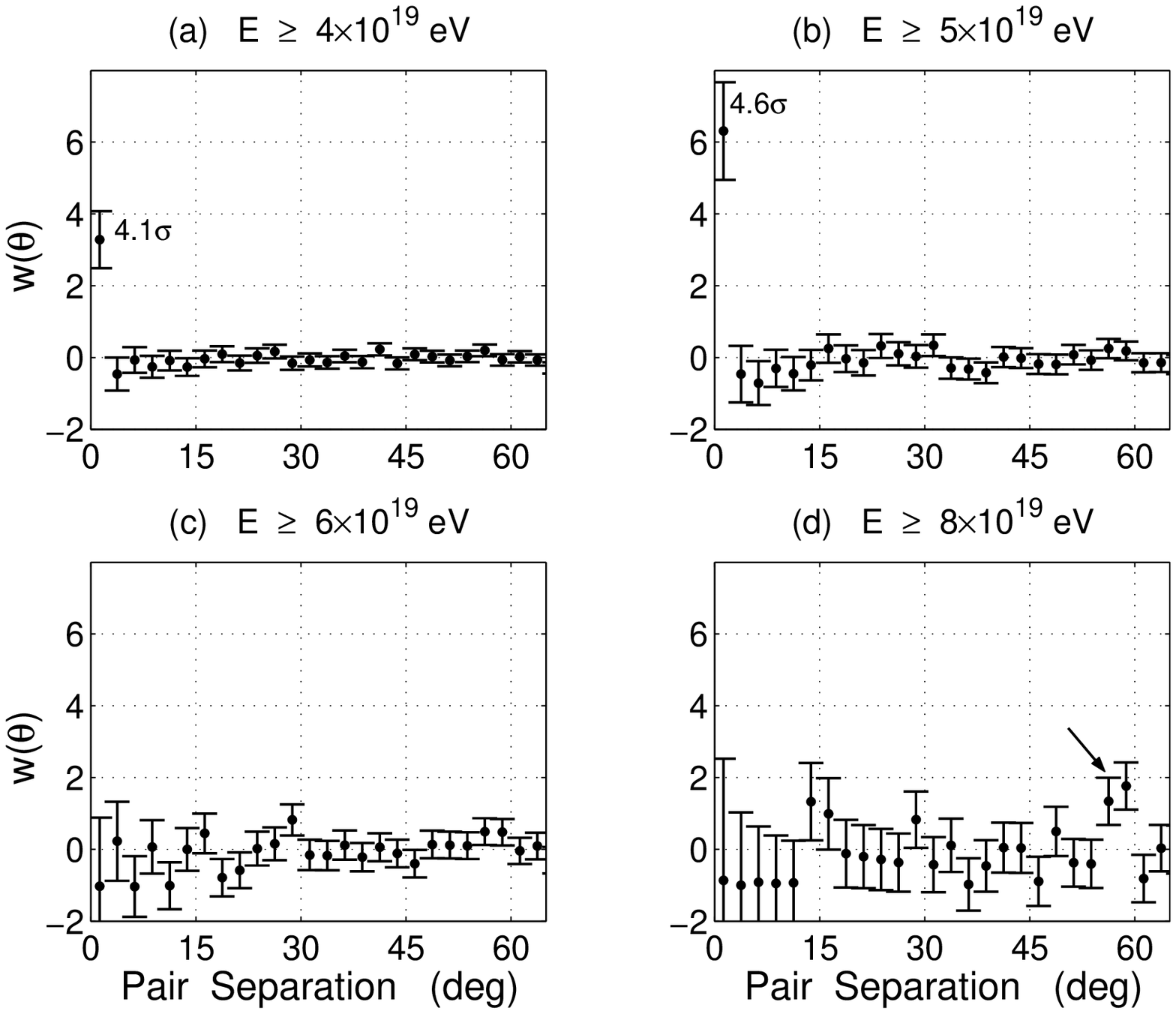}
\caption{Arrival directions of UHECRs of energy $E>4\times10^{10}\gev$
(circles) and $E>10^{11}\gev$ (squares) , with the `doublets' and
`triplet' highlighted (the shaded area is unobservable by AGASA)
\protect\cite{Takeda:1999sg}. The bottom panel shows the angular
autocorrelation function for events above various energy thresholds
\protect\cite{Burgett:2003yg}.}
\label{agasaaniso}
\end{figure}

The other important new result concerns the AGASA data on the angular
distribution of events on the sky \cite{Takeda:1999sg}. As shown in
Fig.~\ref{agasaaniso}, although the large-scale distribution of the 59
observed showers with $E>E_\GZK$ is consistent with isotropy, there
are a number of `clusters' --- defined as a grouping of 2 or more
events within (approximately the experimental angular resolution of)
2.5$^0$. The chance probability that the 5 observed doublets and 1
triplet result from an isotropic distribution is estimated by Monte
Carlo to be less than $10^{-4}$ \cite{Takeda:1999sg}. However
Fig.~\ref{agasaaniso} shows that when the events are partitioned by
energy, small-scale clustering is seen only for events with
$E<6\times10^{10}\gev$; at higher energies there is {\em no} clustering
\cite{Burgett:2003yg}.

\begin{figure}[htb]
\centering
\includegraphics[width=0.68\textwidth]{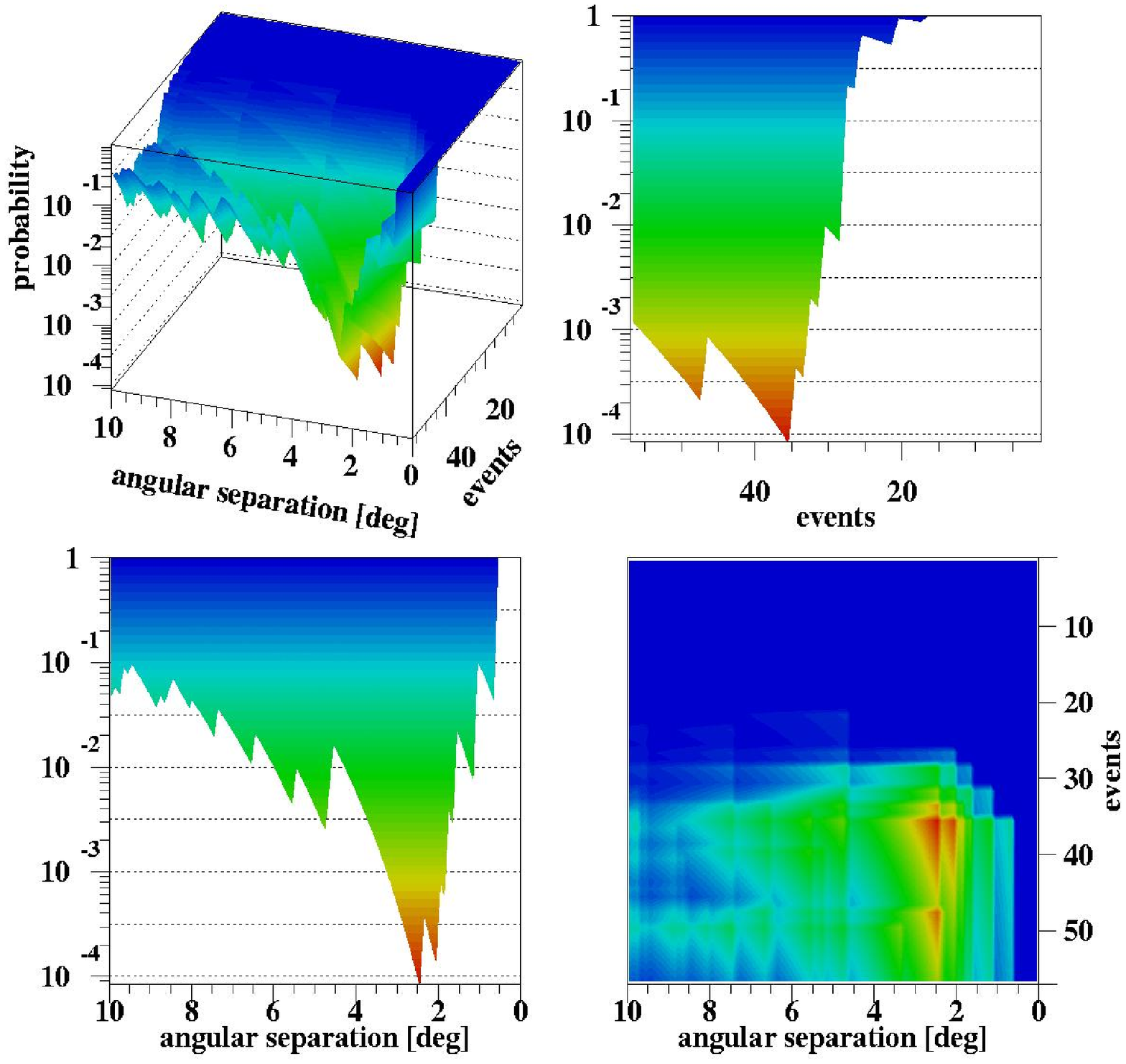}\\
\includegraphics[width=0.7\textwidth]{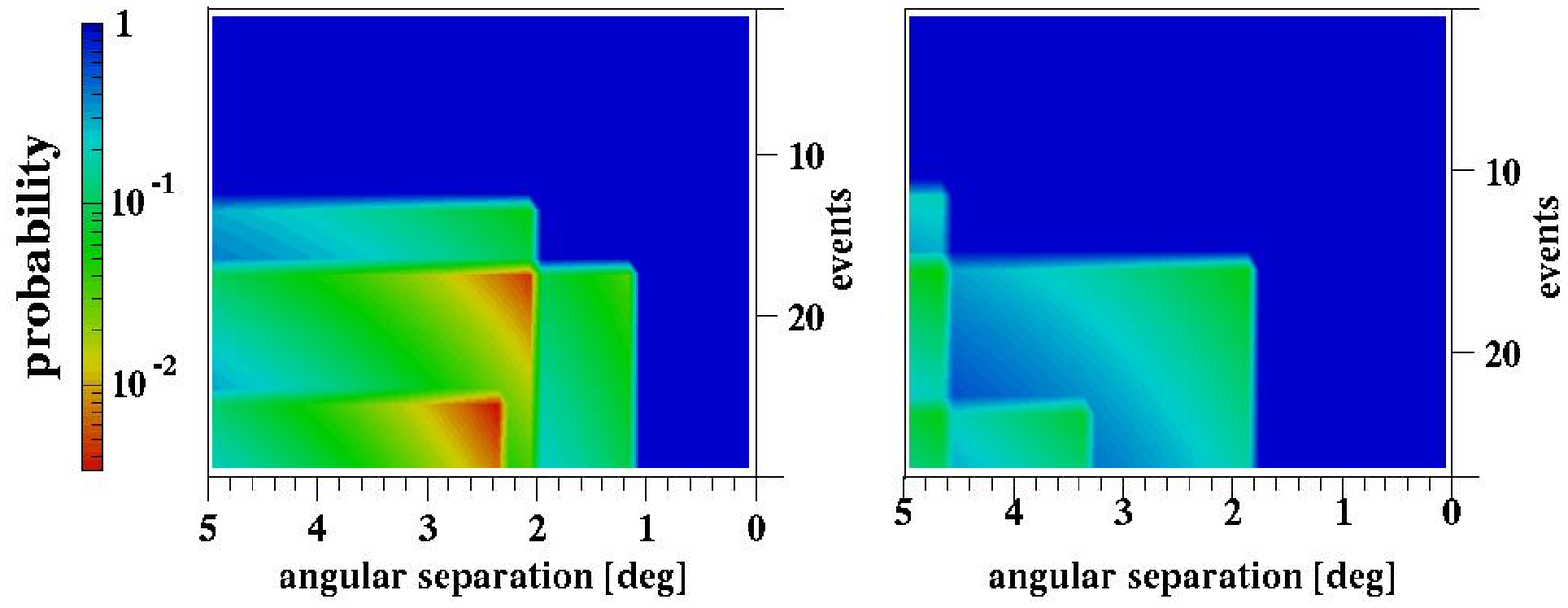}
\caption{Autocorrelation scan of 57 AGASA events with
$E>4\times10^{10}\gev$ shown in 4 views
\protect\cite{Finley:2003ur}. The bottom panel shows the chance
probabilities for the observed clustering separately for the 30 and 27
events, observed before and after Oct 1995.}
\label{agasascan}
\end{figure}

To estimate the significance of the clustering reliably, one should
calculate the `penalty factor' for making {\em a posteriori} cuts (in
energy thresholds and angular separations) on the data set in order to
maximize the clustering signal. Secondly, as was emphasized earlier
\cite{Watson:2001et}, the data set used to make the initial claim for
clustering ought not to be used in the actual analysis. Recently both
these issues have been investigated in detail \cite{Finley:2003ur}. As
shown in Fig.~\ref{agasascan}, the strongest autocorrelation signal
($P^{\data}_{\min}\simeq8.4\times10^{-5}$) is seen at separation angle
$\theta_\c=2.5^0$, with energy threshold
$E_\c=4.9\times10^{10}\gev$. However Monte Carlo tests show that 3475
out of $10^6$ simulated AGASA data sets have
$P^{\MC}_{\min}\leq\,P^{\data}_{\min}$ so the chance probability for
this is 0.35\%. This is 10 times higher than the value claimd earlier
\cite{Tinyakov:2001ic} --- these authors did not allow for their
(arbitrary) choice of $\theta_\c$ to maximise the clustering signal
and interpreted their result as implying that the {\em ``Correlation
function of ultra-high energy cosmic rays favors point
sources''}. Secondly, when the AGASA data is divided into 2 roughly
equal sets (30 events detected before October 1995 and 27 events
detected afterwards), the chance probabilities jump to 4.4\%
($\theta_\c=2.4^0$) and 27\% ($\theta_\c=4.7^0$) respectively for the
observed clustering \cite{Finley:2003ur}! The reason appears to be
that the two sub-sets of data are strongly correlated --- the
`triplet' being made of 2 events from the first set and 1 from the
second. Moreover AGASA has announced more events bringing to 42 the
number detected since October 1995. There are 2 `doublets' separated
by less than $2.5^0$ in this data set but the chance probability for
having observed these is 19\% \cite{Finley:2003ur}. Thus the
clustering observed by AGASA is {\em not} statistically significant,
so does not require the observed UHECRs to originate from point
sources.

\begin{figure}[htb]
\centering
\includegraphics[width=0.4\textwidth]{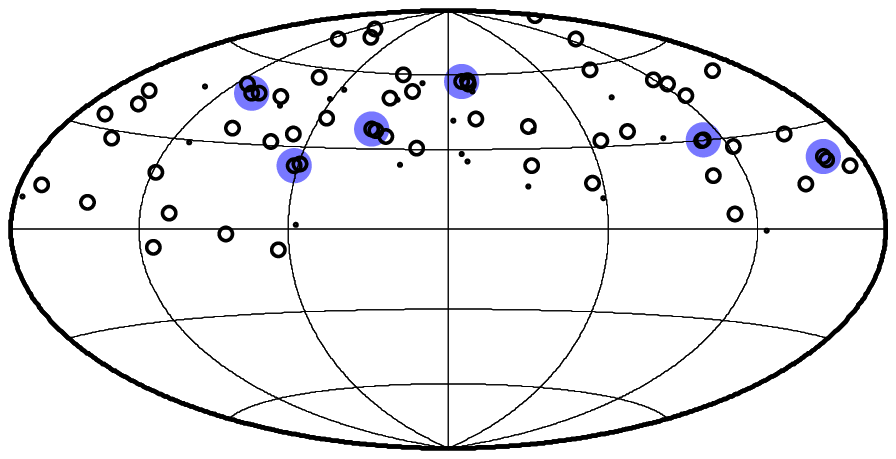}
\includegraphics[width=0.4\textwidth]{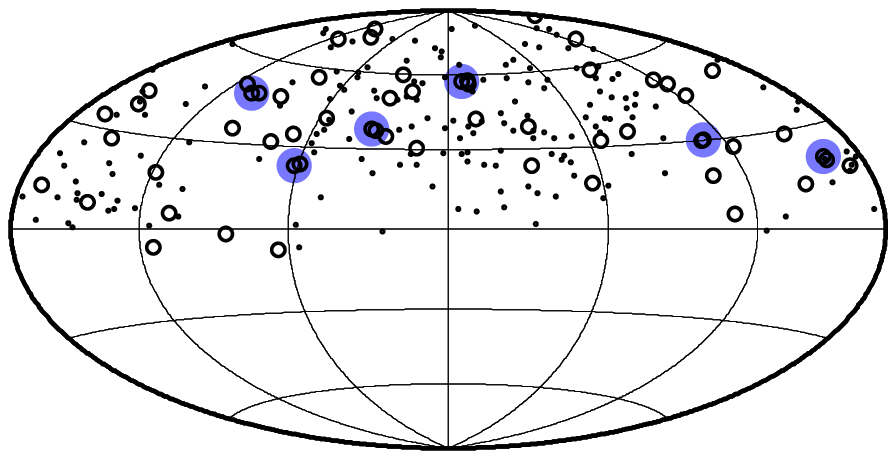}
\caption{The sky distribution of 57 UHECRs (circles) with
$E>4\times10^{10}\gev$ observed by AGASA, with the 5 `doublets' and 1
`triplet' highlighted. The left panel shows the 22 BL Lacs (dots)
satisfying specific cuts on redshift, magnitude and 6~cm radio flux
\protect\cite{Tinyakov:2001nr}, while the right panel shows all 306 BL
Lacs in the catalogue \cite{Evans:2002ry}.}
\label{bllacs}
\end{figure}

Nevertheless, some authors have sought for and found statistially
significant correlations between UHECR arrival directions and possible
sources such as (specific subsets of) active galactic nuclei. In
particular a selected sample of 39 AGASA events (with
$E>4.8\times10^{10}\gev$) and 26 Yakutsk events (with
$E>2.4\times10^{10}\gev$) are claimed to have significant alignments
within $2.5^0$ with 22 selected BL Lacartae objects (having redshift
$z>0.1$ or unknown, magnitude $m<18$ and 6 cm radio flux $F_6 > 01.7$
Jy). The value of $P^{\data}_{\min}$ is $4\times10^{-6}$ and the
penalty factor for making cuts on the BL Lac catalogue is estimated
to be only 15, yielding a chance probability of $6\times10^{-5}$
\cite{Tinyakov:2001nr}. However the Yakutsk experiment has an angular
resolution worse than $4^0$ at the low (sub-GZK) energy cut made to
maximise the coincidences, so these cannot be physically meaningful;
dropping these events increases the chance probability for the
remaining coincidences (with AGASA events) to 0.15\%
\cite{Evans:2002ry}. Also the assumption that the {\sl ``energies of
the events are not important for correlations at small angles''}
\cite{Tinyakov:2001nr} is clearly wrong since a mild drop in the
energy threshold for AGASA events to $E>4\times10^{10}\gev$ decreases
the significance further by a factor of 5
\cite{Evans:2002ry}. Relaxing the cuts on the BL Lac catalogue yields
only 2 coincidences between BL Lacs and AGASA `doublets', with a
chance probability of 6.3\% \cite{Evans:2002ry}. Moreover an
independent sample of 27 Haverah Park and 6 Volcano Ranch events with
$E>4\times10^{10}\gev$ do not coincide at all with the chosen 22 BL
Lacs \cite{Torres:2003ee}. Thus, as shown in Fig.~\ref{bllacs}, there
is no justification for the claim that {\sl ``BL Lacertae are sources
of the observed ultra-high energy cosmic rays''}
\cite{Tinyakov:2001nr}.

Finally, the inferred composition of UHECRs (as well as their
energies) is sensitive to the UHE interaction model used
\cite{Knapp:2002vs} and this realization has muddled the picture
suggested earlier by Fly's Eye and HiRes of a change from heavy
nucleus domination at $3\times10^8\gev$ to nucleon domination at
$10^{10}\gev$. Present data is consistent with a mixed composition at
all energies. Reanalysis of `horizontal showers' at Haverah Park
requires that no more than 50\% of UHECRs above $4\times10^{10}\gev$
can be photons \cite{Ave:2001xn} and a similar (but weaker) limit has
been set by AGASA on the basis that photon-initiated showers tend to
be muon-poor \cite{Shinozaki:ve}. However the showering of UHE photons
is rather complex (because of pair conversion in the geomagnetic field
and the LPM effect), e.g. it cannot be excluded that the highest
energy Fly's Eye event ($E\sim3\times10^{11}\gev$) was in fact
initiated by a photon \cite{Homola:2003bm}.

\section{Conventional explanations}

{\em No} astrophysical object has ever been definitively identified
(e.g. through $\gamma$-ray or $\nu$ emission) as an accelerator of
high energy nucleons. However there do exist energetic objects such as
$\gamma$-ray bursts, active galactic nuclei, or extended lobes of
radio galaxies which satisfy the `Hillas criterion' of being big
enough and/or having sufficiently large (estimated) magnetic fields to
be able to confine UHECRs upto $\sim10^{12}\gev$ \cite{Nagano:ve}. The
issue of whether particles can actually be accelerated to the observed
energies in such objects is an open one, in particular considerations
of radiative energy losses during the acceleration process set very
strong constraints
\cite{Aharonian:2002we,Medvedev:2003sx}. Ultrarelativistic bulk flows
in $\gamma$-ray bursts (GRBs) remain a possibility, and this coupled
with the HiRes claim of a GZK cutoff in the energy spectrum, have
revitalized the suggestion that such cosmologically distant objects
are the sources of UHECRs \cite{Bahcall:2002wi,Vietri:2003te}. However
the energetic requirements are formidable --- such sources must inject
$\sim10^{44}$ erg/Mpc$^3$/yr in UHECRs of energy
$10^{10}-10^{12}\gev$ in order to match the observed flux, while GRBs
are observed to emit roughly this much power at far smaller energies
of ${\cal O}(1)$ MeV. It is hard to conceive of a particle
acceleration mechanism that can generate comparable amounts of power
in such widely separated energy regions. Nevertheless if there is
indeed a GZK cutoff, this is probably the most attractive possibility
for the sources of UHECRs.

\section{Models involving new physics}

However if the UHECR spectrum does extend without a GZK cutoff as
indicated by AGASA, then new physics would appear to be required. We
discuss below only those possibilities which have not been ruled out
already. (In particular we do not consider whether the primaries might
be neutrinos having enhanced interaction cross-sections through new
physics such as TeV-scale extra dimensions --- this {\em cannot}
explain the observed UHECRs, although studies of airshowers can in
principle probe such new physics \cite{Anchordoqui:2002hs}. We also
disregard the hypothesis that the primaries are new light stable
hadrons (for which the GZK cutoff energy can be higher), since such particles
are {\em excluded} by laboratory experiments \cite{Hagiwara:fs}.)

\subsection{Violation of Lorentz invariance}

It was noted long ago that a small modification of the relation
between momentum and energy in special relativity may undo the GZK
cutoff \cite{Sato:2003tv}. If the clustering of events claimed by
AGASA is substantiated by the forthcoming high statistics data from
Auger, then point sources of UHECRs would be implicated. In that case
it may be appropriate to invoke such violation of Lorentz invariance
to explain the absence of the GZK cutoff in the spectrum. However it
is hard to see how this possibility can be falsified since the
required violation is so small that it need not manifest itself in any
other astrophysical or laboratory phenomenon \cite{Jacobson:2002hd}.

\subsection{Z-bursts}

Since at least one species of relic neutrinos must have a mass
$\gtrsim0.1\,\ev$, it is attractive to suppose that these provide a
target for UHE neutrinos from distant sources to annihilate on. This
would create `Z-bursts' with an energy of
$m_Z^2/2m_\nu\sim4\times10^{12}(m_\nu/1\,\ev)^{-1}\gev$, i.e. in the
right energy range to be a source for UHECRs
\cite{Weiler:1997sh,Fargion:1997ft}. The energy spectrum of the
nucleons and $\gamma$-rays resulting from $Z$ decays is well known so
a detailed comparison can be made with the AGASA data. Agreement is
obtained for a relic neutrino mass of $0.08-1.3$~eV
\cite{Fodor:2002hy}, however the required UHE $\nu$ flux is very
large, taking into account that such light relic neutrinos cannot
cluster significantly \cite{Singh:2002de}. Although there are no
direct bounds yet at these energies, it seems implausible that the
required extragalactic sources of $\sim10^{13}\gev$ neutrinos can
exist, given the restrictive bounds on deeply penetrating air showers
from experiments such as Fly's Eye, as well as AGASA and RICE
\cite{Anchordoqui:2002vb}. Moreover it is not clear how such high
energy neutrinos can be created in the hypothetical cosmic sources ---
usually this would require the acceleration of even higher energy
nucleons!

\subsection{Decaying supermassive dark matter}

The possible existence of relic metastable massive particles whose
decays can create high energy cosmic rays and neutrinos had been
discussed \cite{Ellis:1990nb,Gondolo:1991rn} before the detection of
the famous Fly's Eye event with $E\sim3\times10^{10}\gev$ which
focussed attention on the possible absence of the GZK
cutoff. Subsequently this idea was revived and it was noted that such
particles, being cold dark matter, would naturally have a overdensity
by a factor of $\sim10^4$ in the halo of our Galaxy
\cite{Berezinsky:1997hy,Birkel:1998nx}. Hence if their slow decays
generate UHECRs, all propagation effects will be unimportant (except
possibly for photons) in determining the observed spectrum and
cosmposition, and, moreover, there should be a detectable anisotropy
in the arrival directions, given our asymmetric position in the
Galaxy. In order to account for the highest energy events with the
observed rates, the particle mass should exceed
$m_X\gtrsim10^{12}\gev$ while to match the UHECR flux its lifetime
must be $\tau_X\simeq3\times10^{19}\zeta_X$~yr, where $\zeta_X$ is the
fraction of the halo dark matter in the form of such particles. It is
conceivable that such particles exist in the `hidden sector' of
string/M-theory \cite{Benakli:1998ut} and can be produced with a
cosmologically interesting abundance at the end of inflation
\cite{Chung:1998ua}.

\begin{figure}[htb]
\centering 
\includegraphics[width=0.7\textwidth]{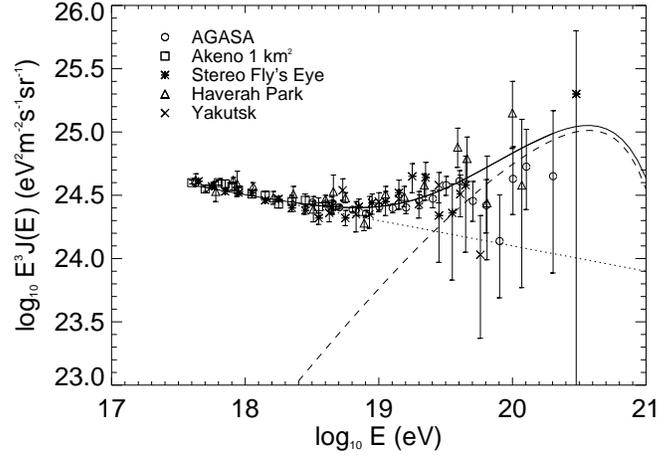} 
\caption{Fit to the UHECR spectrum beyond $E_\GZK$ (excluding the
HiRes data) with a decaying dark matter particle mass of
$5\times10^{12}\gev$ (dashed line); the dotted line is the extension
of the spectrum observed at lower energies
\protect\cite{Sarkar:2001se}.}
\label{fit}
\end{figure}

\begin{figure}[htb]
\centering
\includegraphics[width=0.48\textwidth]{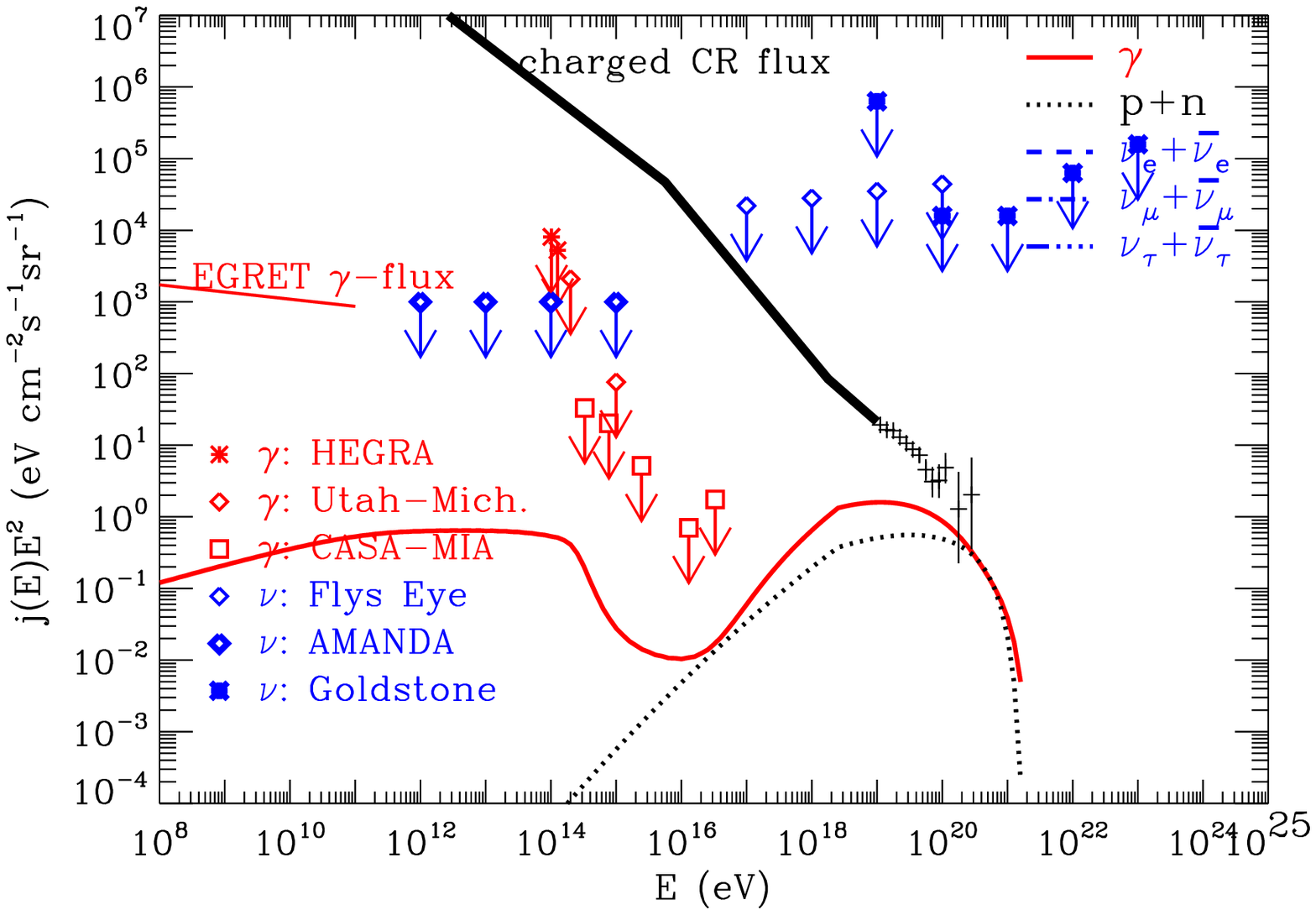}
\includegraphics[width=0.48\textwidth]{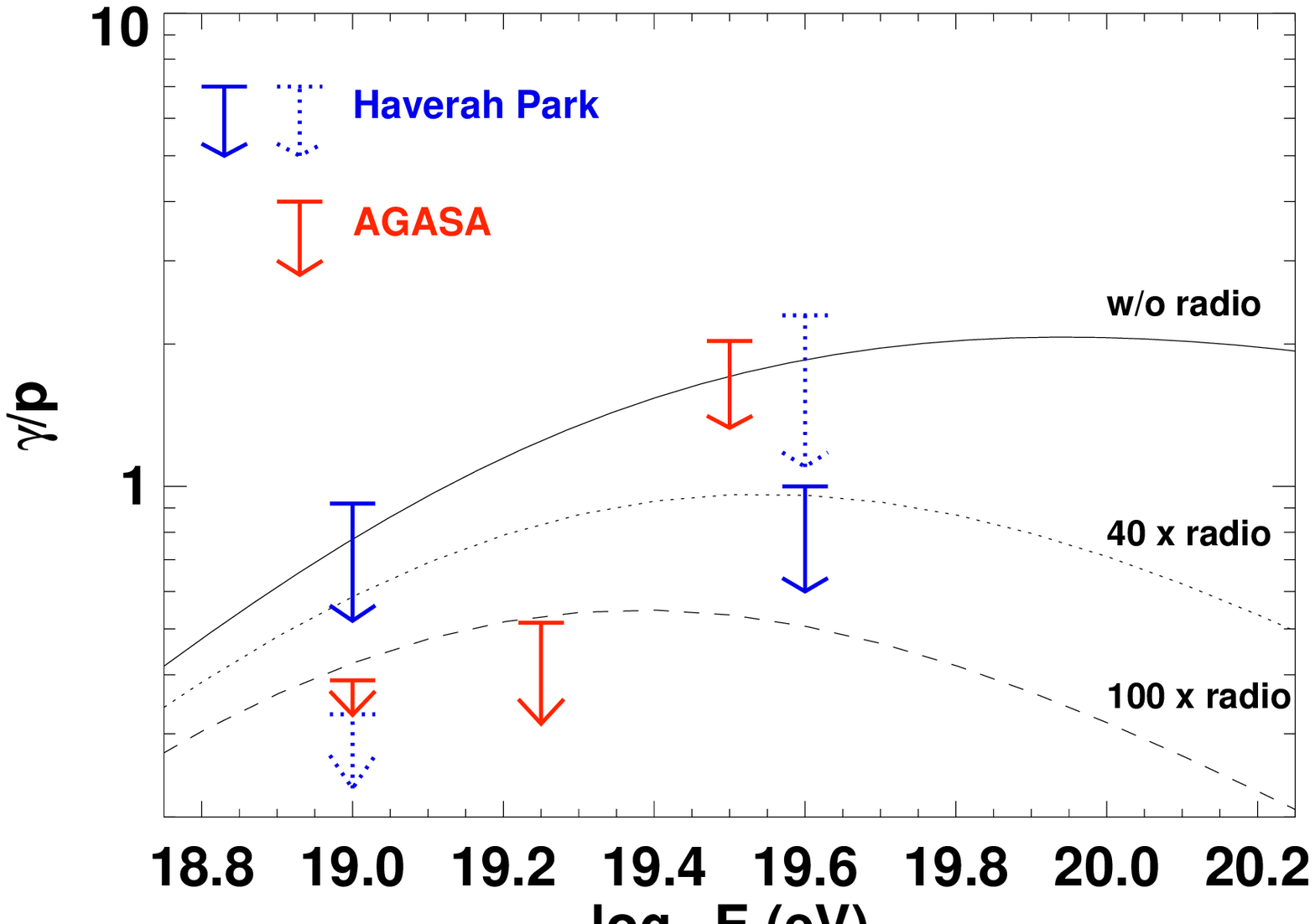}
\caption{The expected processing of the UHE photon spectrum from
decaying dark matter in the Galactic halo is shown (left panel) for a
radio background 10 times higher than is usually assumed and the
$\gamma$/p ratio for various assumed values of the radio background is
compared with experimental limits (right panel)
\protect\cite{Sigl:2002}.}
\label{gamma}
\end{figure}

The spectra of the decay products is essentially determined by the
physics of QCD fragmentation so can be calculated e.g. by evolving the
fragmentation functions measured in $Z^0$ decay upto the energies of
interest using the DGLAP equations
\cite{Rubin:1999,Fodor:2000za,Sarkar:2001se}. Recent work
\cite{Barbot:2002gt} has included electroweak corrections and a more
careful treatment of the possible effects of supersymmetry. As seen in
Fig.~\ref{fit}, the evolved spectrum of nucleons is in good agreement
with the `flat' component of UHECR extending beyond $E_\GZK$. However
the decay photons, which have a similar spectral shape are more
abundant by a factor of $\sim2$, so this model (as well as a similar
subsequent proposal involving annihilations rather than decays
\cite{Blasi:2001hr}) would seem to be ruled out by the experimental
bounds \cite{Ave:2001xn,Shinozaki:ve} on the photon component of
UHECRs. It is possible that the attenuation of such UHE photons in the
halo of the Galaxy (through pair production on $\sim$MHz frequency
radio photons) has been underestimated --- the radio background
intensity assumed for such calculations is very uncertain
\cite{Protheroe:1996si}. However, such attenuation would in turn
generate a background of low energy $\gamma$-rays and it is necessary
to check that this does not exceed observational limits, particularly
from EGRET at $\sim100$~MeV. As shown in Fig.~\ref{gamma}, such
constraints can indeed be satisfied although the required increase in
the intensity of the radio background is rather large
\cite{Sigl:2002}. Perhaps one should wait for Auger to establish
whether photons are indeed ruled out as the UHECRs.

\begin{figure}[htb]
\centering 
\includegraphics[width=0.5\textwidth]{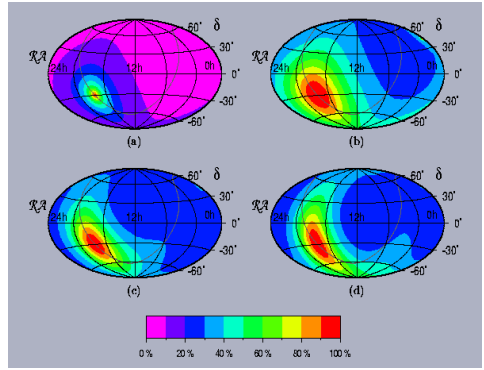}
\caption{Contour plots of the UHECR sky for (a) cusped, (b)
isothermal, (c) triaxial and (d) tilted models of the dark matter halo
of our Galaxy \protect\cite{Evans:2001rv}.}  \label{halo}
\end{figure}

Detailed calculations have also been made of the expected anisotropy,
adopting different possible models of the dark matter halo (cusped,
isothermal, triaxial and tilted) \cite{Evans:2001rv}. The amplitude of
the anisotropy is controlled by the extent of the halo, while the
phase is controlled by its shape. As seen in Fig.~\ref{halo}, the
amplitude of the first harmonic is $\sim0.5$ for a cusped `NFW' halo
but falls to $\sim0.3$ for an isothermal halo which is more favoured
by observations \cite{Binney:2001wu}, while the maximum is in the
direction of the Galactic Centre with deviations of up to $30^0$ for
triaxial and tilted haloes. To detect the predicted anisotropy (for
the likely case of an isothermal halo with large core radius) will
require detection of $\sim500$ UHECRs by Auger; this should also
suffice to determine the angular phase to within $\pm20^0$
\cite{Evans:2001rv}.

Another signature of this hypothesis is the expected UHE neutrino
flux, which was indeed the first to be studied
\cite{Gondolo:1991rn}. Recent detailed calculations
\cite{Barbot:2002kh} indicate that the expected flux should yield
$\sim10-40$ events/yr with $E>10^5\gev$ in IceCube \cite{icecube} and
a similar number of events in RICE \cite{rice}. Of course other
proposed models for UHECRs also predict associated fluxes of UHE
neutrinos. However an unique test is the expected flux of neutralinos
if these are the lightest stable supersymmetric particles
\cite{Berezinsky:1997sb}. In principle EUSO \cite{euso} can
discriminate such events from neutrino-induced ones
\cite{Barbot:2002et}.

\section{Conclusions}
After a long hiatus, high energy cosmic rays have again become very
interesting for particle physicists looking for evidence of physics
beyond the Standard Model. The source of the highest energy particles
in Nature is an equally interesting enigma for astrophysicists. As
Lema\^{\i}tre first suggested, the origin of such particles may even
be linked to the early universe, although not quite as he imagined!
Presently the data are tantalising but not sufficient in either
quantity or quality to distinguish definitively between proposed
models. The good news that this will soon be remedied by the Pierre
Auger Observatory \cite{Blumer:2003bc}. About 100 water Cerenkov
detectors and 2 fluoresence detectors are now operational, covering an
area twice that of AGASA, and the full array should be operational by
the end of 2005. Moreover the ambitious space-based experiment EUSO
\cite{euso}, scheduled for a 3-year engineering flight on the
International Space Station `Columbus' in 2007, will provide another
substantial increase in collecting power. This is an exciting time for
cosmic ray physics and we look forward to the surprises that Nature
has in store for us.

\newpage
\section{Acknowledgements} It is a particular pleasure to present this
talk at Krak\'ow which has had such a distinguished history of cosmic
ray research --- I understand that the first international IUPAP
conference on cosmic rays was in fact held here in 1947, during which
Powell announced the discovery of the pion! I wish to thank Paolo
Lipari and Henryk Wilczynski for their kind invitation, and all the
organisers of this meeting for their efficiency and hospitality.


\begin{thebibliography}{99}

\bibitem{Nagano:ve}
M.~Nagano and A.~A.~Watson,
Rev.\ Mod.\ Phys.\  {\bf 72}, 689 (2000).

\bibitem{Anchordoqui:2002hs}
L.~Anchordoqui, T.~Paul, S.~Reucroft and J.~Swain,
Int.\ J.\ Mod.\ Phys.\ A {\bf 18}, 2229 (2003).

\bibitem{Takeda:2002at}
M.~Takeda {\it et al.} [AGASA collab.],
Astropart.\ Phys.\  {\bf 19}, 447 (2003).

\bibitem{Abu-Zayyad:2002sf}
T.~Abu-Zayyad {\it et al.} [HiRes collab.],
arXiv:astro-ph/0208301.

\bibitem{Bergman:2002pw}
D.~R.~Bergman [HiRes collab.],
Nucl.\ Phys.\ B, Proc.\ Suppl.\ {\bf 117}, 106 (2003);
Proc. 28th Intern. Cosmic Ray Conference, Tsukuba, Vol.\ 1, p.\ 683 (2003).

\bibitem{Blumer:2003bc}
J.~Blumer [AUGER Collaboration],
Proc. 28th Intern. Cosmic Ray Conference, Tsukuba, Vol.\ 1, p.\ 445
(2003); Pierre Auger Observatory homepage: {\tt http://www.auger.org/}

\bibitem{Takeda:1999sg}
M.~Takeda {\it et al.} [AGASA collab.],
Astrophys.\ J.\ {\bf 522}, 225 (1999);
M.~Teshima {\it et al.}, 
Proc. 28th Intern. Cosmic Ray Conference, Tsukuba, Vol.\ 1, p.\ 437 (2003);
AGASA homepage: {\tt
http://www-akeno.icrr.u-tokyo.ac.jp/AGASA}

\bibitem{Burgett:2003yg}
W.~S.~Burgett and M.~R.~O'Malley,
Phys.\ Rev.\ D {\bf 67}, 092002 (2003).

\bibitem{Watson:2001et}
A.~A.~Watson,
arXiv:astro-ph/0112474.

\bibitem{Finley:2003ur}
C.~B.~Finley and S.~Westerhoff,
arXiv:astro-ph/0309159.

\bibitem{Tinyakov:2001ic}
P.~G.~Tinyakov and I.~I.~Tkachev,
JETP Lett.\  {\bf 74}, 1 (2001).

\bibitem{Tinyakov:2001nr}
P.~G.~Tinyakov and I.~I.~Tkachev,
JETP Lett.\  {\bf 74}, 445 (2001).

\bibitem{Evans:2002ry}
N.~W.~Evans, F.~Ferrer and S.~Sarkar,
Phys.\ Rev.\ D {\bf 67}, 103005 (2003).

\bibitem{Torres:2003ee}
D.~F.~Torres, S.~Reucroft, O.~Reimer and L.~A.~Anchordoqui,
Astrophys.\ J.\  {\bf 595}, L13 (2003).

\bibitem{Knapp:2002vs}
J.~Knapp, D.~Heck, S.~J.~Sciutto, M.~T.~Dova and M.~Risse,
Astropart.\ Phys.\  {\bf 19}, 77 (2003).

\bibitem{Ave:2001xn}
M.~Ave, J.~A.~Hinton, R.~A.~Vazquez, A.~A.~Watson and E.~Zas,
Phys.\ Rev.\ D {\bf 65}, 063007 (2002).

\bibitem{Shinozaki:ve}
K.~Shinozaki {\it et al.},
Astrophys.\ J.\  {\bf 571}, L117 (2002).

\bibitem{Homola:2003bm}
P.~Homola {\it et al.},
Proc. 28th Intern. Cosmic Ray Conference, Tsukuba, Vol.\ 1, p.\ 547 (2003).

\bibitem{Aharonian:2002we}
F.~A.~Aharonian, A.~A.~Belyanin, E.~V.~Derishev and V.~V.~Kocharovsky,
Phys.\ Rev.\ D {\bf 66}, 023005 (2002).

\bibitem{Medvedev:2003sx}
M.~V.~Medvedev,
Phys.\ Rev.\ E {\bf 67}, 045401 (2003).

\bibitem{Bahcall:2002wi}
J.~N.~Bahcall and E.~Waxman,
Phys.\ Lett.\ B {\bf 556}, 1 (2003).

\bibitem{Vietri:2003te}
M.~Vietri, D.~De Marco and D.~Guetta,
Astrophys.\ J.\  {\bf 592}, 378 (2003).

\bibitem{Hagiwara:fs}
K.~Hagiwara {\it et al.}  [Particle Data Group Collaboration],
Phys.\ Rev.\ D {\bf 66}, 010001 (2002);
2003 update on: {\tt http://pdg.lbl.gov/}

\bibitem{Sato:2003tv}
H.~Sato,
arXiv:astro-ph/0311306.

\bibitem{Jacobson:2002hd}
T.~Jacobson, S.~Liberati and D.~Mattingly,
Phys.\ Rev.\ D {\bf 67}, 124011 (2003).
 
\bibitem{Weiler:1997sh}
T.~J.~Weiler,
Astropart.\ Phys.\  {\bf 11}, 303 (1999).

\bibitem{Fargion:1997ft}
D.~Fargion, B.~Mele and A.~Salis,
rays,''
Astrophys.\ J.\  {\bf 517}, 725 (1999).

\bibitem{Fodor:2002hy}
Z.~Fodor, S.~D.~Katz and A.~Ringwald,
JHEP {\bf 0206}, 046 (2002).

\bibitem{Singh:2002de}
S.~Singh and C.~P.~Ma,
Phys.\ Rev.\ D {\bf 67}, 023506 (2003).

\bibitem{Anchordoqui:2002vb}
L.~A.~Anchordoqui, J.~L.~Feng, H.~Goldberg and A.~D.~Shapere,
Phys.\ Rev.\ D {\bf 66}, 103002 (2002).

\bibitem{Ellis:1990nb}
J.~R.~Ellis, G.~B.~Gelmini, J.~L.~Lopez, D.~V.~Nanopoulos and S.~Sarkar,
Nucl.\ Phys.\ B {\bf 373}, 399 (1992).

\bibitem{Gondolo:1991rn}
P.~Gondolo, G.~Gelmini and S.~Sarkar,
Nucl.\ Phys.\ B {\bf 392}, 111 (1993).

\bibitem{Berezinsky:1997hy}
V.~Berezinsky, M.~Kachelriess and A.~Vilenkin,
Phys.\ Rev.\ Lett.\  {\bf 79}, 4302 (1997).

\bibitem{Birkel:1998nx}
M.~Birkel and S.~Sarkar,
Astropart.\ Phys.\  {\bf 9}, 297 (1998).

\bibitem{Benakli:1998ut}
K.~Benakli, J.~R.~Ellis and D.~V.~Nanopoulos,
Phys.\ Rev.\ D {\bf 59}, 047301 (1999).

\bibitem{Chung:1998ua}
D.~J.~H.~Chung, E.~W.~Kolb and A.~Riotto,
Phys.\ Rev.\ Lett.\  {\bf 81}, 4048 (1998).

\bibitem{Rubin:1999}
N.~Rubin,
M. Phil. Thesis, University of Cambridge (1999). 

\bibitem{Fodor:2000za}
Z.~Fodor and S.~D.~Katz,
Phys.\ Rev.\ Lett.\  {\bf 86}, 3224 (2001).

\bibitem{Sarkar:2001se}
S.~Sarkar and R.~Toldra,
Nucl.\ Phys.\ B {\bf 621}, 495 (2002).

\bibitem{Barbot:2002gt}
C.~Barbot and M.~Drees,
Astropart.\ Phys.\  {\bf 20}, 5 (2003).

\bibitem{Blasi:2001hr}
P.~Blasi, R.~Dick and E.~W.~Kolb,
Astropart.\ Phys.\  {\bf 18}, 57 (2002).

\bibitem{Protheroe:1996si}
R.~J.~Protheroe and P.~L.~Biermann,
Astropart.\ Phys.\  {\bf 6}, 45 (1996)
[Erratum-ibid.\  {\bf 7}, 181 (1997)].

\bibitem{Sigl:2002}
G.~Sigl, S.~Sarkar and R.~Toldra,
unpublished (2002).

\bibitem{Evans:2001rv}
N.~W.~Evans, F.~Ferrer and S.~Sarkar,
Astropart.\ Phys.\  {\bf 17}, 319 (2002).

\bibitem{Binney:2001wu}
J.~J.~Binney and N.~W.~Evans,
Mon.\ Not.\ R.\ Astron.\ Soc. {\bf 327}, L27 (2001).

\bibitem{Barbot:2002kh}
C.~Barbot, M.~Drees, F.~Halzen and D.~Hooper,
Phys.\ Lett.\ B {\bf 555}, 22 (2003).

\bibitem{Berezinsky:1997sb}
V.~Berezinsky and M.~Kachelriess,
Phys.\ Lett.\ B {\bf 422}, 163 (1998).

\bibitem{Barbot:2002et}
C.~Barbot, M.~Drees, F.~Halzen and D.~Hooper,
Phys.\ Lett.\ B {\bf 563}, 132 (2003).

\bibitem{icecube}
IceCube homepage: {\tt http://icecube.wisc.edu/}

\bibitem{rice}
Radio Ice Cerenkov Experiment homepage: {\tt http://kuhep4.phsx.ukans.edu/~iceman/index.html}

\bibitem{euso}
Extreme Universe Space Observatory homepage: 
{\tt http://www.euso-mission.org/}

\end{thebibliography}
\end{document}